\documentclass[twocolumn,prb,showpacs]{revtex4}
\usepackage{graphicx}
\usepackage{amssymb}
\usepackage{bm}

\newcommand{\GTO}{$\rm Gd_2Ti_2O_7$}
\newcommand{\GSO}{$\rm Gd_2Sn_2O_7$}

\begin{document}
\title{Spin-liquid dynamics of pyrochlore magnets \GTO\ and \GSO}

\author{S.\,S.\,Sosin, L.\,A.\,Prozorova, A.\,I.\,Smirnov}
\affiliation{P.\,L.\,Kapitza Institute for Physical Problems RAS,
119334 Moscow, Russia}
\author{P.\,Bonville}
\affiliation{Commissariat \`{a} l'\'{E}nergie Atomique, Centre de Saclay,
DSM/SPEC, 91191 Gif sur Yvette, France}
\author{G.\,Jasmin - Le Bras}
\affiliation{Commissariat \`{a} l'\'{E}nergie Atomique, Centre de Saclay,
DSM/SPEC, 91191 Gif sur Yvette, France}
\author{O.\,A.\,Petrenko}
\affiliation{Department of Physics, University of Warwick,
Coventry CV4 7AL, UK}

\date{\today}

\begin{abstract}
The spin-liquid phase of two highly frustrated pyrochlore magnets
\GTO\ and \GSO\ is probed using electron spin resonance in the
temperature range 1.3~--~30~K. The deviation of the absorption
line from the paramagnetic position $\nu =\gamma H$ observed in
both compounds below the Curie-Weiss temperature $\Theta_{\rm
CW}\simeq 10$~K, suggests an opening up of a gap in the excitation
spectra. On cooling to 1.3~K (which is above the ordering
transition $T_{N}\simeq 1.0$~K) the resonance spectrum is
transformed into a wide band of excitations with the gap amounting
to $\Delta\simeq 26$~GHz (1.2~K) in \GTO\ and 18~GHz (0.8~K) in
\GSO. The gaps increase linearly with the external magnetic field.
For \GTO\ this branch co-exists with an additional nearly
paramagnetic line absent in \GSO. These low lying excitations with
gaps, which are preformed in the spin-liquid state, may be
interpreted as collective spin modes split by the single-ion
anisotropy.
\end{abstract}

\maketitle

\section{Introduction}
The properties of a Heisenberg antiferromagnet on a pyrochlore
lattice are determined to a great extent by the high degree of
frustration of the nearest-neighbor exchange interaction. The
minimum of the exchange energy can be achieved in an infinite
number of degenerate states with different spin configurations. A
classical consideration of the fluctuations between the
states~\cite{moessner} and quantum calculations of a spin
correlation function~\cite{canals} show that the system should
remain in a collective paramagnetic state with short-range spin
correlations at any non-zero temperature. In real systems, the
interplay of weaker contributions such as the dipolar interaction,
single-ion anisotropy and next-nearest-neighbor exchange can lift
the macroscopic degeneracy of the ground state driving the system
into a particular ordered state. This principle reveals itself as
a delay in the magnetic ordering transition to temperatures well
below the Curie-Weiss temperature. This type of behavior is
demonstrated by two related compounds \GTO\ and \GSO\ with spin
$S=7/2$ and zero orbital momentum $L=0$ for the Gd$^{3+}$ ions.
Both systems have a Curie-Weiss temperature of about 10~K while
the magnetic ordering develops only below
1.0~K.\cite{raju,ramirez,bonville}

The strong degeneracy of the ground state implies the presence of
a macroscopic number (half of the number of magnetic ions for a
pyrochlore lattice) of local soft modes which should acquire the
gaps and dispersion due to weak interactions. Intensive
theoretical studies were performed for various combinations of
interactions.\cite{palmer,gingras,cepas,wills} For example, the
complex $4k$-structure described by $k=(1/2,1/2,1/2)$ and
equivalent wave-vectors which was proposed for \GTO\ by Stewart
{\it et al.,}~\cite{stewart} can be stabilized under the combined
influence of the dipolar interaction and one of the next-nearest
neighbor exchanges. A large number of nonfreezing degrees of
freedom in the correlated state immediately leads to an enhanced
magnetocaloric effect in the vicinity of the saturation field,
where the system undergoes a transition into a single
spin-polarized state and all the modes acquire a Zeeman
gap.\cite{misha} This effect was seen in an adiabatic
demagnetization experiment.~\cite{sosin} However, a comprehensive
study of this type of excitations has yet to be performed. Since
inelastic neutron scattering experiments in these compounds are
hindered by the large absorption cross-section of Gd nuclei, the
electron-spin resonance technique remains the most suitable
microscopic probe of these excitations.

Our recent works on the magnetic resonance in \GTO\
(Ref.~\onlinecite{sosin1}) and in \GSO\ (Ref.~\onlinecite{sosin2})
were mainly devoted to the study of the ordered states, while the
spin-liquid phase has not been investigated in details. At low
temperature, the systems were shown to have a regular resonance
spectra, where some of the resonance lines could be identified and
described theoretically as quasi-local modes. An early high
temperature ESR work~\cite{hassan} reported an unusual resonance
behavior in \GTO, suggesting it is highly anisotropic (contrary to
other experiments). In addition, the influence of the crystal
field was studied by electron paramagnetic resonance of Gd$^{3+}$
ions placed into the nonmagnetic $\rm Y_2Ti_2O_7$ and $\rm
Y_2Sn_2O_7$ matrices.\cite{glazkov,glazkov1} The energy of the
effective single ion anisotropy of the form $D\hat{S}_z^2$,
arising from mixing with excited $L\neq 0$ Gd$^{3+}$ levels,
amounts to $D=0.2$~K for the titanate and 0.14~K for the stannate,
which is comparable to the nearest-neighbor exchange integral
$J\simeq 0.3$~K, but is still far beyond the values necessary to
interpret the anisotropic effects described in
Ref.~\onlinecite{hassan}. This discrepancy served as an additional
motivation for the present ESR study.

This paper describes in detail the resonance properties of \GTO\
and \GSO\ in the spin-liquid phase (above $T_{N}\simeq 1.0$~K). We
have studied several single crystal and powder samples over a wide
frequency range. The distinctive gap in the excitation spectrum of
the ordered phases is shown to develop above the ordering
transition with the value $\Delta\approx 1.2$~K and 0.8~K for
\GTO\ and \GSO\ respectively. This gap increases as a linear
function of the external field independently of its orientation
with respect to the crystal axes. We also found a new type of
excitations coupled with the microwave field polarized along the
external magnetic field. The strong orientational dependences
previously reported in Ref.~\onlinecite{hassan} are attributed to
electrodynamic resonances in the finite size samples with large
magnetic permeability.

\section{Experimental Details and Results}
\subsection{Sample preparation and experimental techniques} \label{exp}
The single crystal of \GTO\ was grown by the floating zone
technique described in Ref.~\onlinecite{oleg}. Samples of the same
series were previously used to obtain the magnetic phase
diagram~\cite{oleg1} and in the investigation of the enhanced
magnetocaloric effect,~\cite{sosin} as well as for previous low
temperature magnetic resonance studies.~\cite{sosin1} Two
plate-shaped samples of different sizes were cut from the original
single crystal: a large sample (\#1) with the dimensions $1\times
1\times 0.2$~mm$^3$ (about 1~mg), and a small sample (\#2) of
approximately $0.5\times 0.5\times 0.1$~mm$^3$ (0.15~mg). The
samples were cut so that the plane of the plate coincided with the
(111) plane of the crystal lattice. The preparation procedure for
powder samples is briefly outlined in Ref.~\onlinecite{bonville}.
A small amount of powder (0.2~--~0.3~mg) was taken for each
measurement.

We performed a series of ESR experiments in the frequency range 9
to 140~GHz using three wideband transmission-type spectrometers
with rectangular and cylindrical cavities of different sizes.
Samples were glued inside the cavity either on one of its walls or
(for single crystal plates) onto a small worm-wheel used to rotate
the sample from outside the spectrometer. The cavity was inside
the vacuum cell filled with Helium heat exchange gas, immersed
into liquid $^4{\rm He}$ and supplied with a heater and
thermometer allowing a regulation of the temperature between 4.2
and 100~K. Temperatures below 4.2~K (down to the minimum
experimental temperature 1.3~K) were achieved and regulated by
pumping the helium bath. A magnetic field of up to 80~kOe was
generated by a cryomagnet. The resonance spectra were obtained by
recording the transmitted signal on the forward and backward field
sweeps. The microwave power transmitted through the cavity with a
sample inside is determined by the relation:

\begin{equation}
P=P_0\left ( 1+2\pi\chi^{\prime\prime}\alpha Q \right )^{-2},
\label{transmit}
\end{equation}
where $\chi^{\prime\prime}$ is the dynamic susceptibility of the
sample, $\alpha$ is a filling factor, $Q$ is a quality factor of
the cavity, the latter two parameters being different for each
cavity resonance mode. The dynamic susceptibility
$\chi^{\prime\prime}$ may correspond to one or several resonance
lines with Lorentzian shape in angular velocity $\omega$. In case
of a linear relation between $\omega$ and the applied magnetic
field, $H$, each of the lines can be expressed in the following
form:

\begin{equation}\label{lorentz}
\chi^{\prime\prime} =\frac{A}{\left ( 1+\frac{(H-H_{\rm
res})^2}{\Delta H^2} \right )} + \frac{A}{\left (1+\frac{(H+H_{\rm
res})^2}{\Delta H^2} \right )},
\end{equation}
where $H_{\rm res}$ and $\Delta H$ are the position and the
half-linewidth of the line respectively, $(H\mp H_{\rm res})$
originates from the decomposition of the linear polarization of a
microwave field into counterclockwise and clockwise circular
polarizations. Using expressions (\ref{transmit}) and
(\ref{lorentz}), one can fit the experimental absorption signals
by the sets of parameters $W=\alpha QA$ (the signal amplitude),
$H_{\rm res}$ and $\Delta H$. In the case of a single Lorentzian,
the curve can be integrated with respect to the field in order to
approximately relate these parameters to the static susceptibility
$W\Delta H/H_{\rm res}=\alpha Q\chi_{\rm st}$. Firstly, this
allows the temperature dependence of the static susceptibility to
be obtained from the EPR data at a given frequency and secondly,
it provides a reference for comparing signals obtained at
different cavity modes with uncertain $\alpha$ and $Q$.

\subsection{Magnetization of \GTO\ and \GSO}
\begin{figure}[tb]
\centerline{\includegraphics[width=\columnwidth]{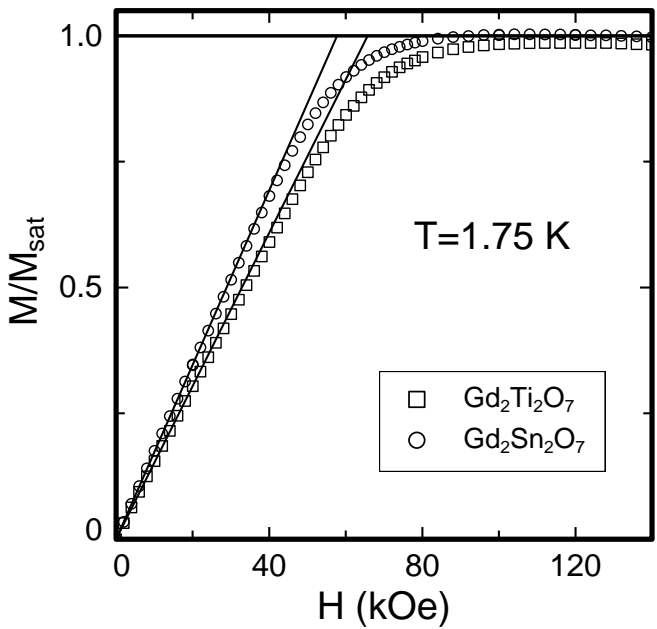}}
\caption{Magnetization curves at $T=1.75$~K scaled to the
saturation magnetization $M_{\rm sat}=g\mu_BS$ in \GTO\ ($\Box$)
and \GSO\ ({\Large $\circ$}) powder samples; solid lines are
linear fits of the low field parts of the curves.} \label{fig0}
\end{figure}

The starting point of the experiment was to test the magnetization
process in powder samples of both compounds under fields up to
140~kOe exceeding by a factor of 2 the transition into the
saturated state. The isothermal magnetization curves were measured
at $T=1.75~K$ in a commercial VSM magnetometer. They were carried
out in a step mode, with a field increase rate of 5~kOe/min. The
sample thermalization is provided by a continuous He gas flow. No
hysteresis on the backward field sweep was observed. The
saturation moment was found to be exactly $7.0~\mu_B$ per Gd ion
in \GSO\ and $6.9~\mu_B$ per Gd ion in \GTO. The magnetization
curves are presented in Fig.~\ref{fig0}, with their values scaled
to the theoretical saturation moment $M_{\rm sat}=g\mu_BS=7\mu_B$.
The initial part of both curves are linear in a rather wide field
range (the corresponding fits are shown in Fig.~\ref{fig0} by
solid lines), which allows one to estimate the nearest-neighbour
exchange constants. In a molecular field approximation, the
relation $M/M_{\rm sat}=g\mu_BH/8JS$ gives the values of
$J=0.32$~K and 0.28~K respectively for \GTO\ and \GSO. The
saturation field can be also estimated by extrapolating these
linear fits to $M=M_{\rm sat}$ which yields $H_{\rm sat}=66$~kOe
and 58~kOe. The result for \GTO\ exceeds by 10\% the previously
reported value of 59-61~kOe measured by ESR in the ordered phase
for different orientations of the magnetic field.\cite{sosin1} A
saturation field of 60~kOe was also obtained from pulse field
magnetization measurements at $T=0.3$~K,\cite{narumi} so our
extrapolation procedure with the data obtained at the temperature
above the ordering transition gives an overestimate for the values
of $H_{\rm sat}$ as well as for the exchange constants $J$.

One should mention that because of the large susceptibility, a
correction to the internal magnetic field due to demagnetization
effects in \GTO\ single crystal samples amounts to 10\% for the
field oriented perpendicular to the sample plates. Hereafter, we
shall present the magnetic field as the internal field, i.e the
one corrected for the demagnetization effect (using the above
curves as well as data taken from Ref.~\onlinecite{bonville}).

\subsection{Electron spin resonance spectrum of \GTO}
\label{esr}

\begin{figure}[tb]
\centerline{\includegraphics[width=\columnwidth]{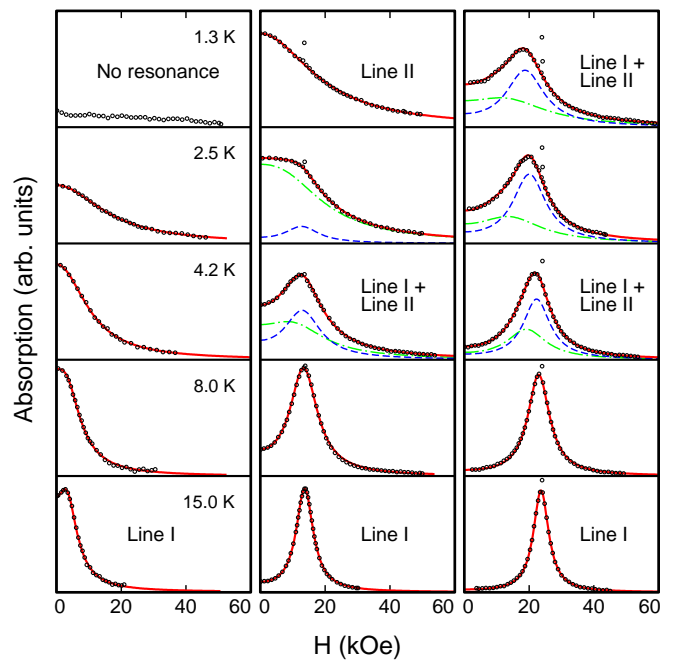}}
\caption{(color online). The field dependence of the absorption
in \GTO\ single crystals for $H\parallel [111]$ at the frequencies
$\nu= 9.6$~GHz (sample~\#1, left panel), $\nu= 36.1$~GHz
(sample~\#1, middle panel) and $\nu =69.6$~GHz (sample~\#2, right
panel). Solid lines are two-component Lorentzian fits as described
in the text, the dashed and dashed-dotted lines show the Line~I
and Line~II of the spectrum separately; out-of-curve points are
DPPH labels used as a reference.} \label{fig1}
\end{figure}

\begin{figure}[tb]
\centerline{\includegraphics[width=\columnwidth]{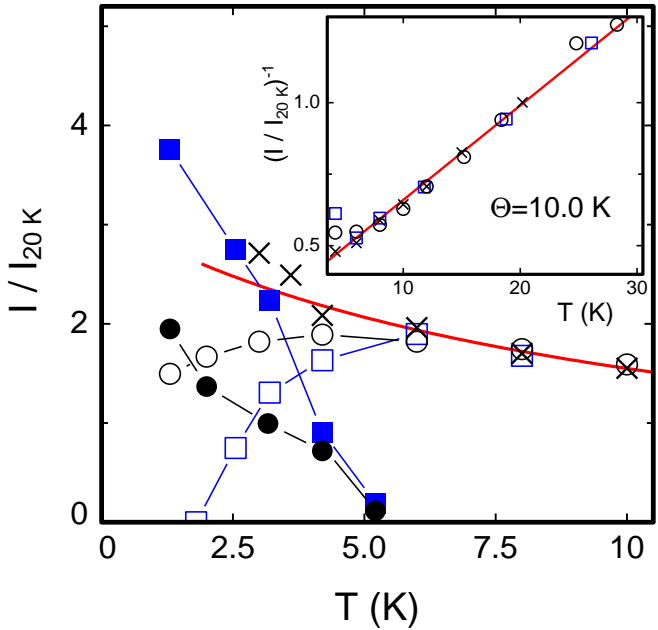}}
\caption{(color online). The temperature dependence of the
integrated intensity $I=W\Delta H$ of two resonance lines
determined at various frequencies from fitting the transmitted
signal by a single or double Lorentzian form (as described in the
text) and scaled to the value obtained for 20~K; $\times$ -- $\nu
=9.6$~GHz (Line~I), $\square$ and $\blacksquare$ -- 36.1~GHz,
{\Large $\circ$} and {\Large $\bullet$} -- 69.6~GHz (respectively
Lines I and II). The inset represents the inverse values of the
integrated intensity in the whole temperature range, the solid
line is a paramagnetic fit with the Curie-Weiss temperature
$\Theta =10$~K.} \label{fig4}
\end{figure}

In the main part of the present work we studied the temperature
evolution of the resonance absorption in \GTO\ and \GSO\ samples
at various microwave frequencies. The experimental results for the
magnetic field applied parallel to $[111]$ axis of a single
crystal sample of \GTO\ are summarized in Fig.~\ref{fig1}. The
left, middle and right panels show the absorption curves recorded
at the lowest, intermediate and high frequencies of 9.6~GHz,
36.1~GHz and 69.6~GHz respectively. At temperatures above the
Curie-Weiss temperature ($T>\Theta =10$ K) a single resonance line
is observed over the whole frequency range. The best fits of these
curves using a single Lorentzian form~(\ref{lorentz}) are shown in
Fig.~\ref{fig1} by the solid lines. The corresponding fitting
parameters $H_{\rm res}$, $\Delta H$ and $W\Delta H$ are analyzed
in Figures~\ref{fig4},\ref{fig5}. The resonance fields at all
frequencies correspond to a paramagnetic $g$-factor equal to 2.0
for all orientations of the magnetic field (the dashed line on the
upper panel of Fig.~\ref{fig5}). This picture remains
qualitatively the same at intermediate temperatures $T\sim\Theta$,
with all the lines broadened and amplified. This behavior is
typical for a concentrated paramagnet at high temperature, where
the different spectral lines (split due to anisotropic effects)
are narrowed into a single line due to the exchange interaction.

On decreasing the temperature further, the resonance lines
continue to broaden and start to shift to lower fields. At the
lowest frequency $\nu =9.6$~GHz the line moves to zero field by
$T=4.2$~K and then decreases in amplitude and disappears
completely at $T=1.3$~K (see left panel of Fig.~\ref{fig1}). This
indicates that no observable (excited by the microwave field)
resonance modes with energies below 10~GHz are left in the
vicinity of the ordering transition at $T\geq T_{N1}=1.0$~K. For
frequencies below 30~GHz this tendency is preserved so that one
can trace the gradual shift of the resonance line towards zero
field on cooling until some temperature after which it is replaced
by a broad line with a maximum absorption at $H=0$.
Fig.~\ref{fig4} shows the integrated intensity scaled to the
intensity at $T=20$~K. Down to approximately 2.5--3.0~K it obeys
the Curie-Weiss law $\chi \sim 1/(T+\Theta)$ with $\Theta =10$~K
(solid line on Fig.~\ref{fig4}) in accordance with previous
magnetization measurements.~\cite{gingras,bonville}

\begin{figure}[tb]
\centerline{\includegraphics[width=\columnwidth]{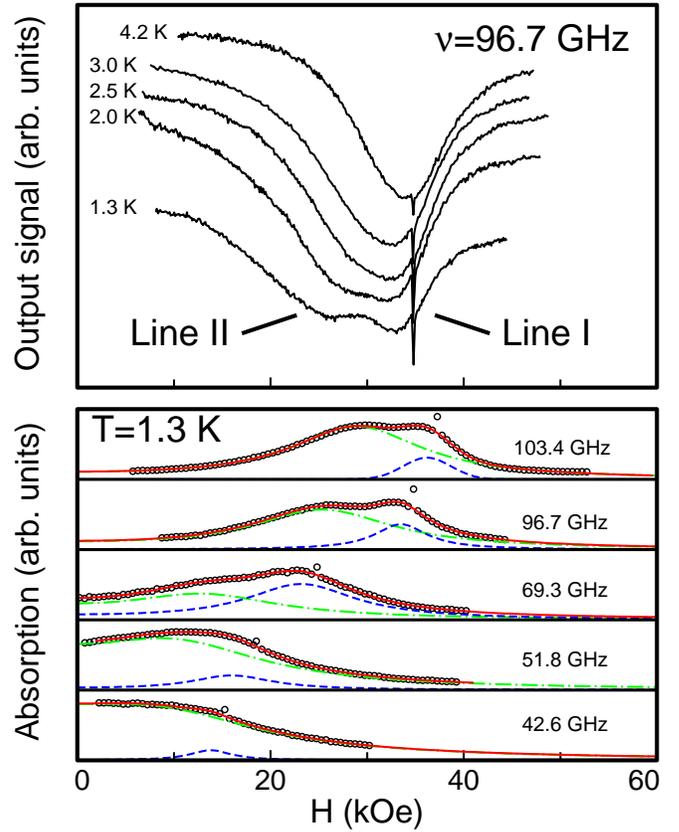}}
\caption{(color online). The field dependence of the resonance
absorption in \GTO\ powder sample for various temperatures at $\nu
=96.7$~GHz (upper panel) and at different frequencies at $T=1.3$~K
(lower panel); solid, dashed and dashed-dotted lines are the same
as in Fig.~\protect\ref{fig1}, narrow signal and out-of-curve
points are DPPH labels.} \label{fig2}
\end{figure}

A different behavior is observed at higher frequencies of 36.1 and
69.6~GHz. Below 4-6~K the resonance spectra cannot be fitted by a
single Lorentzian line. An improvement is achieved by adding a
second absorption, also with Lorentzian shape (the middle and
right panels of Fig.~\ref{fig1}). One of these lines remains
around the paramagnetic position (Line~I) while the second one
shifts to smaller fields similarly to the previously described
behavior at lower frequencies (Line~II). At $\nu =36.1$~GHz the
intensity of the Line~I rapidly decreases (as shown by open
squares in the Fig.~\ref{fig4}) while the intensity of the Line~II
increases (closed squares) and finally becomes the only resonance
response of the sample at $T=1.3$~K. This response has properties
that are significantly different from those of the paramagnetic
signal. It is also observed at the frequency $\nu =69.6$~GHz,
although the intensity of the paramagnetic line does not drop to
zero at the minimum temperature. Because of the large width of
both lines at low temperatures ($\Delta H\simeq 8-10$~kOe and
15-20~kOe for the Line~I and II respectively at $T=1.3$~K), the
dependence of the signal on the magnetic field direction cannot be
traced.

On further increasing the measurement frequency the absorption of
the single crystal samples is disrupted by a parasitic
electrodynamic effect (described in details below in Subsection
\ref{anis}) and the measurements are only possible on powder
samples. The typical temperature evolution of the signal from the
powder sample obtained at $\nu =96.7$~GHz is presented in the
upper panel of Fig.~\ref{fig2}. The single line positioned nearly
to the paramagnetic DPPH label observed at $T=4.2$~K is replaced
by a double signal, with Line~II increasing in intensity and
shifting to smaller fields at lower temperatures. Since the
linewidth of both signals tends to decrease at high frequencies
(at larger values of the resonance fields) the splitting of the
resonance absorption into two components becomes more prominent
and easier to trace. The comparison of signal records taken at
different frequencies (the lower panel of Fig.~\ref{fig2})
illustrates this tendency and provides unambiguous proof of the
resonance line doubling in \GTO\ at temperatures below 4~K.

\subsection{Electron spin resonance spectrum of \GSO}
\label{esr1}

\begin{figure}[tb]
\centerline{\includegraphics[width=\columnwidth]{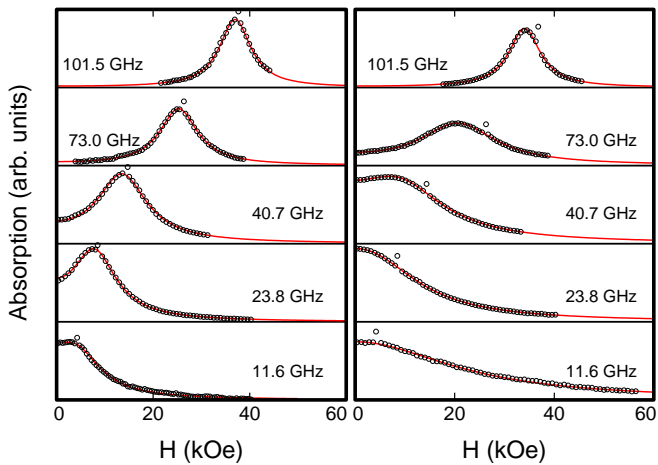}}
\caption{(color online). The field dependence of the resonance
absorption in \GSO\ powder sample for various frequencies at
$T=4.2$~K (left panel) and 1.3~K (right panel); solid lines are
single Lorentzian fits to the experimental points.} \label{fig3}
\end{figure}

The absorption spectra of \GSO\ are presented in Fig.~\ref{fig3}
for two temperatures: 4.2~K (in the left panel) and the minimum
experimental temperature 1.3~K (right panel). A single resonance
line observed at 4.2~K was found to gradually broaden and shift to
smaller fields with no trace of splitting. The splitting is
evidently absent even above 100~GHz, where the linewidth is small
enough ($\Delta H\leq 5$~kOe) for the two components to be reliably
resolved as it is the case for \GTO\ (see the previous Section).

\subsection{Longitudinal microwave susceptibility} \label{long}
The results of the previous section were obtained for the
microwave field $h_{\rm mw}$ polarized perpendicular to the
external magnetic field. Such a polarization is typical for ESR
experiments. The high frequency response of \GTO\ sample \#2 was
also studied in a ``parallel'' polarization of the microwave field
for which the spin precession of the paramagnetic type cannot be
excited. Instead, one can probe the longitudinal oscillations of
the magnetic moment associated with the internal degrees of
freedom of the correlated spin system. The microwave polarization
for the $TE_{01n}$ modes in the rectangular cavity can be selected
by placing the sample either on the vertical centerline of the
wide cavity wall ($h_{\rm mw}\perp H$) or on its narrow wall
($h_{\rm mw}\parallel H$). In both cases the polarization is
almost perfect as the sample size is much smaller than the
dimensions of the cavity.

\begin{figure}[tb]
\centerline{\includegraphics[width=\columnwidth]{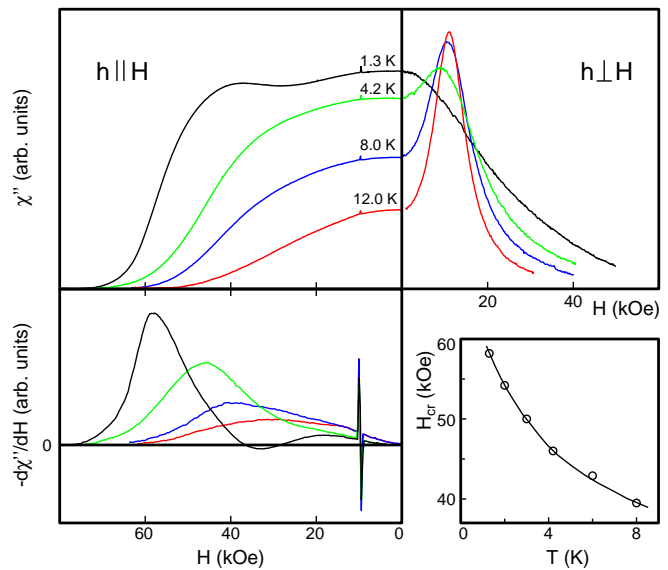}}
\caption{(color online). Field dependence of the resonance
absorption and its derivative (upper and lower left panels
respectively) at the frequency $\nu= 27.1$~GHz for the microwave
field polarized along the direction of the external magnetic
field, and the corresponding data for the perpendicular microwave
field polarization (upper right panel). Lower right corner inset
shows the temperature evolution of the crossover into the ground
state with destroyed ``soft modes"; the solid line is a guide to
the eye.} \label{fig6}
\end{figure}

A comparison between the sample responses observed at a frequency
$\nu =27.1$~GHz for the two polarizations of the microwave field
is presented in Fig.~\ref{fig6}. The data for $h_{\rm mw}\parallel
H$ and $h_{\rm mw}\perp H$ are rescaled to each other by the
values of $\chi^{\prime\prime}(H=0)$ obtained at $T=12$~K. The
field dependences of the sample absorption at different
temperatures for $h_{\rm mw}\parallel H$ are shown in the upper
left panel. The high temperature curve ($T=12$~K) at low fields
shows an absorption with an amplitude corresponding to the wing of
a Lorentzian resonance line of the form~(\ref{lorentz}). The
intensity of this signal is attenuated at fields comparable to the
saturation field of an $S=7/2$ paramagnet. The absorption for the
perpendicular polarization shown on the upper right panel
corresponds to a paramagnetic line with $g=2.0$ that is well
fitted by~(\ref{lorentz}). On lowering the temperature the
magnetic field required to suppress the longitudinal
susceptibility of the system increases reaching 60~kOe at
$T=1.3$~K. This value exceeds by a factor of 10 the paramagnetic
saturation field and corresponds to the transition of the exchange
correlated structure into a fully polarized state. Roughly the
same value 66~kOe was obtained from our magnetization
measurements. The shift of the suppression field is illustrated on
the lower left panel by the derivative of the absorption by
magnetic field. The temperature dependence of the corresponding
inflection points is shown on the lower right inset to
Fig.~\ref{fig6}. For the perpendicular polarization (see the upper
right panel) the perfect paramagnetic line developed at high
temperatures is gradually transformed into a broad absorption with
the maximum positioned at zero field (as described in previous
sections). By comparing the $T=1.3$~K curves for $h_{\rm
mw}\parallel H$ and $h_{\rm mw}\perp H$ one can conclude that the
strongly correlated disordered spin state in \GTO\ possesses low
energy internal degrees of freedom amenable to the longitudinal
microwave field. The absorption intensity is reduced by an
external field applied perpendicular to the direction of the
microwave field but is practically insensitive to the parallel
magnetic field up to the crossover into the state where the
collective degrees of freedom are destroyed by temperature and
magnetic field. In the low temperature limit this crossover
corresponds to the transition into the spin-polarized state in
which the system cannot respond to the weak perturbation of the
saturated magnetic moment.

\subsection{Resonance ``size effects''}\label{anis}
\begin{figure}[tb]
\centerline{\includegraphics[width=\columnwidth]{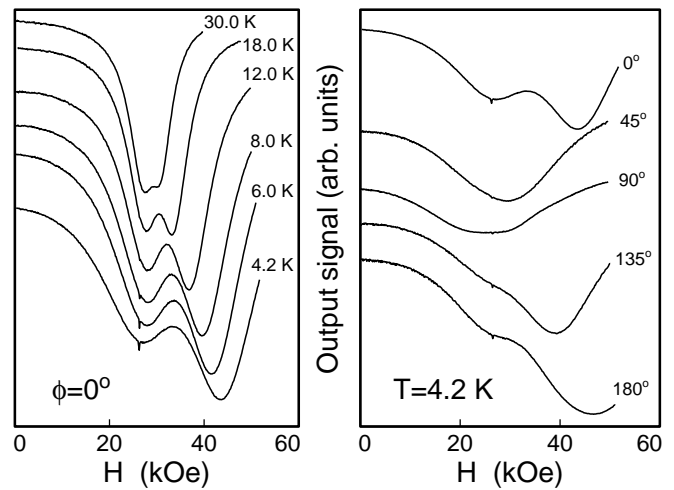}}
\caption{The field dependences of the absorption at $\nu =
69.4$~GHz in the sample \#1 recorded at different temperatures for
$H\parallel [111]$  (left panel), and at $T=4.2$~K for different
sample orientations (right panel); narrow peaks at all curves is a
paramagnetic DPPH label.} \label{fig7}
\end{figure}

The main features of the resonance spectra observed in \GTO\
single crystals are described in Sections~\ref{esr} and
\ref{long}. Here we draw the attention to the unusual resonance
behavior first reported in Ref.~\onlinecite{hassan}, which is also
observed in our samples. At frequencies above 60~GHz the resonance
line in the single crystal sample \#1 appears to be split into two
broad components, both having strong temperature and orientational
dependences whose typical shape is shown in Fig.~\ref{fig7} for
$\nu =69.4$~GHz. The temperature evolution of the absorption
persists over the whole temperature range until 30~K. The signal
also varied strongly with a period of nearly $180^{\circ}$ when
rotating the sample inside the resonator (see Fig.~\ref{fig7},
left panel). Such properties are usually indicative of systems
with a strong uniaxial anisotropy, with a value much larger than
the exchange interaction; this was indeed suggested in
Ref.~\onlinecite{hassan} to explain the results presented there.
Nevertheless, this suggestion obviously contradicts all the other
experimental data on the compound, which show the single-ion
anisotropy constant does not exceed 0.2~K (see for example
Ref.~\onlinecite{glazkov}).

\begin{figure}[tb]
\centerline{\includegraphics[width=\columnwidth]{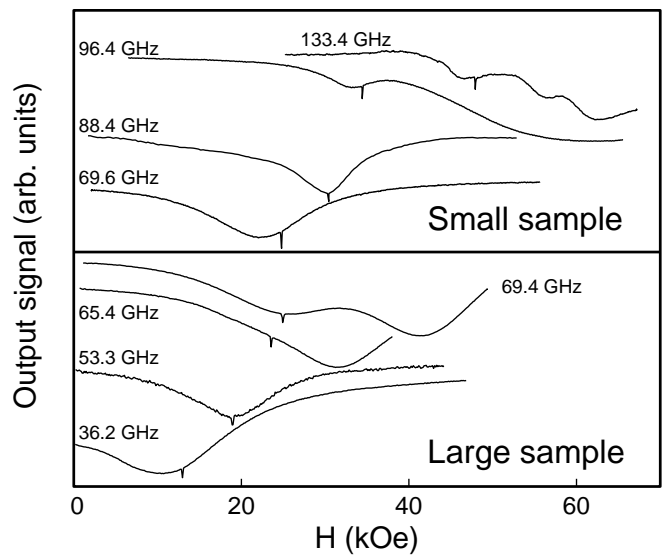}}
\caption{The comparison of absorption spectra of the sample \#1
(lower panel) and sample \#2 (upper panel) taken at $T=4.2$ K for
several frequencies; the upper curve for the large sample and the
lower curve for the small sample are obtained at the same
frequency of the cylindrical cavity.} \label{fig8}
\end{figure}

Our experiments show this problem can be eliminated by decreasing
the size of the sample. Fig. \ref{fig8} illustrates how the
absorption signal is modified for sample \#2 with linear
dimensions of approximately one half of sample \#1. The resonance
absorption, measured at the same cavity frequency in order to
avoid any influence of polarization effects, demonstrates dramatic
changes in the response to the microwave field. Namely, the
intense high field absorption with unusual temperature and
orientational properties disappears and the resonance line
acquires the shape described in the Section~\ref{esr} above (for
comparison, see the corresponding curves on the lower and upper
panels of Fig.~\ref{fig8} taken at $\nu =69.4$ and 69.6~GHz
respectively). On further increasing the frequency of the
measurement the splitting reappears (see the curve at $\nu
=96.4$~GHz) revealing the same characteristic features as a
function of temperature and sample orientation.

We therefore suggest that this effect is associated not with the
properties of a pyrochlore, but rather with the geometry of the
sample, its large dielectric constant $\varepsilon$ and its field
dependent permeability $\mu$. When one of the linear sample
dimensions $l$ roughly meets the condition

\begin{equation}
l \sim n \frac{\lambda} {2\sqrt{\varepsilon\mu}}, \label{res}
\end{equation}

\noindent where $\lambda$ is the electromagnetic wavelength in
free space and $n$ is an integer, an electrodynamic resonance
takes place, with the sample playing the role of the resonator,
and its absorption rises. The smaller the sample size $l$, the
higher the boundary frequency for this effect to reveal itself.
The value of $\mu^{\prime} =1+4\pi\chi^{\prime}$ is enhanced in
\GTO\ as a result of the large spin value of Gd and a relatively
small saturation field of the magnetic structure. In addition, the
field dependence of $\chi^{\prime}$ has a singularity in the
vicinity of the paramagnetic resonance field $\chi^{\prime}\sim
(H-H_{\rm pm})/(1+(H-H_{\rm pm})^2/\Delta H^2)$ which can be
responsible for meeting the resonance condition (\ref{res}) twice
(or more) during a field scan. One of these ``resonance'' fields
is positioned near $H_{\rm pm}$ while the other should be shifted
away from it. The value of this shift increases on lowering the
temperature due to the increase of the singularity amplitude as
was observed in our experiment. In the high frequency limit, the
number of such electrodynamic modes grows, resulting in further
splitting of the absorption line (see the 133.4~GHz curve in
Fig.~\ref{fig8}). Due to the coupling with the main cavity this
signal should be sensitive to the orientation of the sample plate
with respect to the distribution of the microwave field in the
cavity, which explains the observed $180^{\circ}$ periodic
dependence. One should mention that the effect described above is
only present for continuous samples. In order to extend the
frequency range over which reliable results can be obtained, we
used powder samples spread over the resonator wall in the shape of
a thin rarefied film.

\section{Discussion and conclusions}
\begin{figure}[ht]
\centerline{\includegraphics[width=\columnwidth]{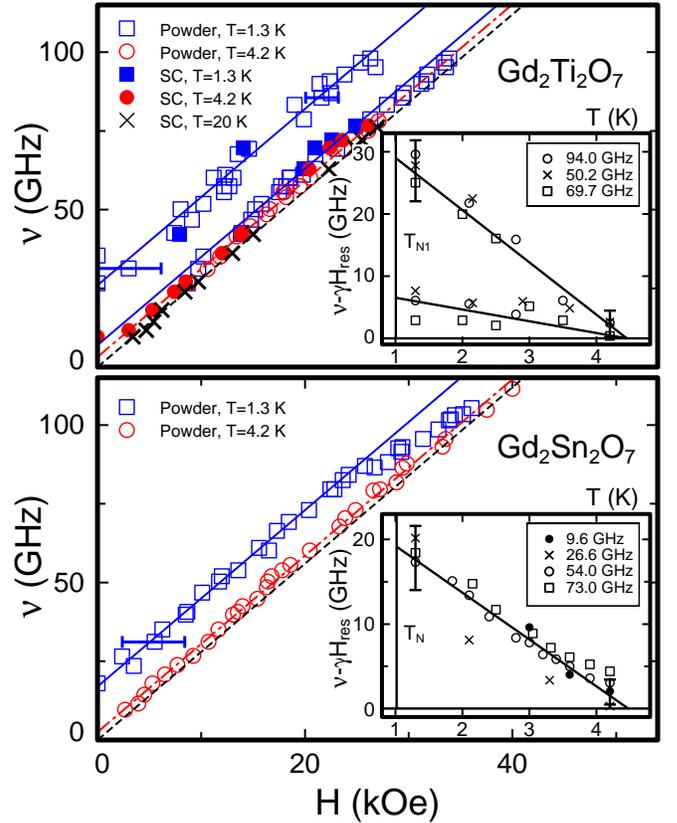}}
\caption{(color online). The frequency-field diagram of the
obtained ESR spectra. Upper panel -- \GTO\ powder sample
($\square$ and {\Large $\circ$}) and single crystal with
$H\parallel [111]$ ($\blacksquare$, {\Large $\bullet$} and
$\times$) at different temperatures. Lines are $\nu =\Delta
+\gamma H$ ($\gamma = g\mu_B/h$ with $g=2.0$) fits to the data
points: dashed, dashed-dotted and solid lines correspond to
$T=20$~K, 4.2~K and 1.3~K respectively with the gaps $\Delta =0$
(a typical concentrated paramagnet), 3.5~GHz, 5 and 26~GHz. The
inset represents the temperature dependence of the value $\nu
-\gamma H_{\rm res}$ for the two resonance modes at
various frequencies. Lower panel -- analogous results for the
powder sample of \GSO\ at $T=4.2$~K ({\Large $\circ$}) and 1.3~K
($\square$) described by gapped linear dependences with $\Delta =
2.5$~GHz (dashed-dotted line) and 18~GHz (solid line); the inset
is the same as for the \GTO\ panel, but with the single resonance line.
(Error bars on both panels correspond to $\Delta H/5$).}
\label{fig5}
\end{figure}

We begin the analysis of the data for \GTO\ and \GSO\ described in
the previous sections starting from the high temperature region
$T\geq \Theta_{\rm CW}$, where a single paramagnetic
resonance line corresponding to an isotropic $g$-factor of 2.0 is
observed in both systems (its frequency-field dependence is shown
by dashed lines in Fig.~\ref{fig5}). This line results from the
exchange narrowing of a set of spectral components split by the
single-ion anisotropy. The initial broadening of this line on
decreasing the temperature down to $T\simeq \Theta_{\rm CW}$ and
its shift to smaller magnetic fields reflect the redistribution of
spin sublevel populations. Such a behavior is common for concentrated
paramagnets.

The single resonance line corresponding to the paramagnetic
precession can be traced in the frequency range 9--100~GHz down to
approximately 4--6~K. The temperature dependence of the intensity
of this signal in \GTO\ single crystal obtained by integrating the
Lorentzian line over a field scan (proportional to the static
susceptibility of the sample) is nicely fitted by a Curie-Weiss
law with $\Theta\approx 10.0$~K (see the inset to Fig.~\ref{fig4})
in agreement with previous susceptibility
measurements.\cite{raju,bonville} The corresponding dependence for
\GSO\ is traced in details only below 4.2~K. The resonance fields,
$H_{\rm res}$, at $T=4.2$~K determined from Lorentzian fits at
various frequencies shift from the paramagnetic positions by
1--2~kOe (as shown in the frequency-field diagram of
Fig.~\ref{fig5}) giving the first evidence of a gap opening.
Within the experimental accuracy, the $\nu (H)$ curves in both
compounds obey the linear law $\nu =\Delta +\gamma H$ ($\gamma
=g\mu_B/h$) with the $g$-factor equal to 2.0 and $\Delta =3.5$~GHz
(0.17~K) and 2.5~GHz (0.12~K) for \GTO\ and \GSO\ respectively
(best fits to the experimental points are plotted by dashed-dotted
lines). The further temperature evolution of the spectrum reveals
the considerable differences between the two systems. Namely, the
resonance absorption in \GSO\ always consists of a single line
gradually shifting to lower fields. At $T=1.3$~K it can be again
described as a linear function in magnetic field with $g=2.0$ and
$\Delta =18$~GHz (solid line in the lower panel of
Fig.~\ref{fig5}).

In contrast, the high temperature resonance line of \GTO\ is
transformed into a double line spectrum below 4~K. The position of
the initial spectral line remains almost unchanged while the
second line moves to lower fields similar to the one observed in
\GSO, so that the splitting between the two components amounts to
7-8~kOe at temperatures close to the ordering transition. As a
result, the magnetic resonance spectrum of \GTO\ at $T=1.3$~K
consists of two lines with the gaps $\Delta_{1,2}\simeq 5$~GHz
(Line~I) and 26~GHz (Line~II). The absorption intensity is
redistributed between them so that the nearly paramagnetic Line~I
which is principal at the high temperatures becomes a secondary
feature in the low temperature range. At frequencies below
35-40~GHz it is completely decayed into the wide band of gapped
excitations corresponding to the Line~II absorption and becomes
unobservable in our measurements. Due to a narrowing of the two
lines under magnetic field they become reliably resolved at
frequencies above 90~GHz. The existence, in the vicinity of an
ordering transition, of a residual mode with the gap value not
exceeding 0.25~K may (i) reflect the partial disorder of the
magnetic structure of \GTO\ below ordering proposed in
Ref.\onlinecite{stewart}; (ii) result in the $C\propto T^2$
behaviour of the specific heat observed in \GTO\ well below the
ordering temperature.\cite{yaouanc}

The temperature evolution of the gaps in both compounds are
illustrated in the insets of Fig.~\ref{fig5}. The represented
parameters $\nu -\gamma H_{\rm res}$ are equivalent to the gap
values on the assumption that the resonance frequency increases
linearly under magnetic field. These parameters start to deviate
from zero at temperatures below 4-5~K and then increase linearly
to the values specified above. The relation between the gaps at
$T=1.3$~K and the single-ion anisotropy constants in both
compounds (anisotropy of the form $D\sum S_z^2$ determined from
EPR measurements \cite{glazkov,glazkov1}) $\Delta /D=1.2/0.21$ and
$0.8/0.14$ are consistent with each other and roughly equal to 6.

Note that this energy exactly corresponds to the transition
$|-7/2\rangle\rightarrow |-5/2\rangle$ (for $D<0$) or
$|5/2\rangle\rightarrow |7/2\rangle$ (for $D>0$) of an isolated
$S=7/2$ magnetic ion when the magnetic field is applied along the
anisotropy axis. Nevertheless, for a concentrated paramagnet at
high temperatures this transition is impossible to observe
separately of the transitions between other levels. As mentioned
above, the resulting line should be narrowed by the exchange
interaction in the vicinity of the paramagnetic position, as was
observed in our experiment. In the spin-liquid phase formed at
temperatures $T\ll\Theta_{CW}$ this description is not applicable,
since the magnetic resonance spectrum should be related to the
excitations above the collective ground state. The influence of
short-range correlations on the magnetic resonance in spin systems
with no long-range ordering in the ground state and gapless
spectrum was first reported for quasi-1D $S=5/2$ spin
chains.\cite{nagata} The shift of the resonance line was
attributed to the combined effect of the dipolar interaction and a
single-ion anisotropy. The origin of a similar effect in
pyrochlore systems is unclear and requires theoretical studies.
Although the crystal field of cubic symmetry should not affect
$S=1$ excitations, the temperature and field evolution of the
observed resonance absorption resembles the case of a system of
triplet levels split by a single-ion anisotropy and magnetic
field. On the other hand, the short-range correlations between
spins on the pyrochlore-like lattice are mostly related to hexagon
loops in $(111)$-planes, \cite{lee} so one can also suppose the
observation in a spin-liquid phase of weakly dispersive modes
gapped due to anisotropy and dipolar interaction. The visual
absence of an orientational dependence of the spectrum might
result from the high symmetry of the crystal lattice. Once the
nature of the collective excitations is known precisely, a
detailed analysis of this problem can be performed.

Other properties of the low temperature excitations that are
significantly different from the conventional paramagnetic
precession appears when the external magnetic field is applied
parallel to the linear microwave field polarization. The
paramagnetic precession cannot be driven in this field
configuration. Therefore, the high temperature response to the
microwave field is observed only at low external fields
corresponding to the wing of the Lorentzian resonance line and
demonstrates the rapid decrease in intensity under high fields
(see Fig.~\ref{fig6}). In contrast, the low temperature resonance
modes can be excited by the longitudinal microwave field and are
therefore associated with the longitudinal oscillations of the net
magnetic moment. These eigen modes of spin oscillations of the
exchange correlated state in the pyrochlore magnet are different
from conventional spin precession. They might be analogous to the
weakly dispersive optical modes of an antiferromagnetic resonance
in the ordered phase. This response remains practically unchanged
over the whole field range up to the saturation field and then
rapidly disappears because the magnetic moment cannot be perturbed
by a weak microwave field in a fully polarized phase.

In summary, the conventional isotropic paramagnetic resonance mode
with $g=2.0$ observed in \GTO\ and \GSO\ at $T\geq\Theta_{\rm
CW}\approx 10$~K gradually transforms into gapped excitations of a
new type at low temperatures. Unlike conventional
antiferromagnets, the gaps develop well above the transition into
an ordered state over a wide temperature interval, with
short-range correlations and no magnetic ordering characteristic
of highly frustrated magnets. These gaps increase linearly with
the magnetic field in a way that is practically independent of the
field orientation with respect to the crystal axes. These
properties might point to the presence of collective spin
excitations split by the single-ion anisotropy.

\section{ACKNOWLEDGMENTS}
The authors thank V~N.~Glazkov and M.~E.~Zhitomirsky for useful
discussions, G.~Balakrishnan for preparing the \GTO\ samples and
M.~R.~Lees for a critical reading of the manuscript. This work is
supported by INTAS YSF 2004-83-3053, RFBR Grant 07-02-00725 and by
RF President Program.

\end{document}